\documentclass[aps,floatfix]{revtex4}

\usepackage{amsmath,bm,amssymb,hyperref,hypcap,ae,tikz,color}
\usepackage{latexsym,textcomp,slashed,epsfig,graphicx,xcolor}

\hypersetup{
	colorlinks=true,
	citecolor=blue,
    linkcolor=magenta,
    urlcolor=blue
	}

\begin{document}
\title{\bf Memory Effect of Gravitational Wave Pulses in PP-Wave Spacetimes}

\author{\bf{Sucheta Datta\footnote{suchetadatta93@gmail.com} and Sarbari Guha\footnote{srayguha@yahoo.com}}}

\affiliation{\bf {Department of Physics, St.Xavier's College (Autonomous), Kolkata 700016, India}}

\maketitle

\section*{Abstract}
In this paper, we study the gravitational memory effect in pp-wave spacetimes due to the passage of a pulse having the form of a ramp profile through this spacetime. We have analyzed the effect of this pulse on the evolution of nearby geodesics, and have determined analytical solutions of the geodesic equations in the Brinkmann coordinates. We have also examined the changes in the separation between a pair of geodesics and their velocity profiles. The separation (along $ x $ or $ y $-direction) increases monotonically from an initial constant value. In contrast, the relative velocity grows from zero and settles to a final non-zero constant value. These resulting changes are retained as memory after the pulse dies out. The nature of this memory is similar to that determined by earlier workers using Gaussian, square, and other pulse profiles, thereby validating the universality of gravitational wave memory.

\bigskip

KEYWORDS: pp-waves, wave pulse, gravitational wave memory

\section{Introduction}

Gravitational waves (GWs) leave behind imprints on the spacetime through which they propagate. The passing gravitational wave imparts a permanent change in the separation between adjacent geodesics, and this permanent change is known as the gravitational wave memory. The presence of memory effect in GW signals can serve as a test of General Relativity.
This has led to several studies of gravitational memory effect \cite{ZELD, BRAG1, SOUR, BRAG2, POLN, NONLIN1, NONLIN2, THOR, FAV, BIERI} in recent times. It is a strong-field effect of gravity that has remained undetected in astrophysical observations till today \cite{OBSV, HUB}. Because of the inherent weakness of GWs, the permanent changes in the displacement and in the time delays in various detector-dependent setup are too small for the available technologies to directly measure them \cite{SUN}.  However, these changes may become observable in the current and planned detectors in the near future
\cite{WISE, AASI, LASKY, NICH2}, provided that their low-frequency sensitivity can be increased further \cite{YU}.
The article \cite{NICH3} explores the possibility of the detection of non-linear memory by the ALIGO and Virgo detectors. The authors in \cite{SUN, GHOSH} have suggested different methods of detection using space-based detectors. Some of the earlier works, such as \cite{BRAG2} and \cite{THOR}, have discussed the experimental prospects for detecting the memories of GW bursts. Thus it is meaningful to assess the nature of memory produced by gravitational waves.

The study of memory effect first appeared in the pioneering works by Zeldovich and Polnarev \cite{ZELD}, and Braginsky and Grishchuk \cite{BRAG1}. These articles have shown that finite duration burst of GWs causes a permanent change in the separation between test particles, which is known as linear memory effect. A non-linear form of memory has been suggested in \cite{NONLIN1, NONLIN2}. However, the authors in \cite{BIERI2} observed both types of effect in linearized gravity. These have later been classified as the ordinary and null memory effects corresponding to massive and massless particles respectively responsible for producing gravitational radiation \cite{TOL1, WINI2}.
Also, a distinction between GW bursts with and without memory has been reported in \cite{BRAG2}, and earlier in the context of the detection of GW bursts \cite{GIBB}.

In a physical system, the GW memory is encoded in the change between the state before and the state after a gravitational wave passes through it. This change can be defined in terms of a net displacement occurring in `freely-falling detectors' due to the passage of GW. This brings about a permanent change in the background Minkowski spacetimes, the one before the arrival of the pulse and the other after its departure, which are not equivalent \cite{IC1, IC2, IC3, IC5}. The velocity of the test particle also exhibits changes after the wave passes \cite{SOUR, BRAG2, POLN, ZHANG1, ZHANG2, IC1, IC2, IC3, IC5}.
In Ref.\cite{ZELD} it is mentioned that the metric perturbation before and after the event changes in case of hyperbolic scattering. The difference between the quadrupole moments of the source of radiation at initial and final times has been referred to as the memory effect in \cite{BRAG1}. The memory of a GW impulse is given by the change in the transverse-traceless part of the Coulomb-type gravitational field arising from the four-momenta of the components of the source \cite{BRAG2}. The wave field $ h_{ij}^{\text{TT}} $ increases from zero, oscillates for a few cycles and finally settles to zero (bursts without memory) or to a non-zero value (bursts with memory). A collision of two or more initially-free masses, or an explosion of a single mass into several free masses moving independently, would give rise to burst GWs with memory \cite{BRAG2, THOR}.
Besides, there are studies on the spin memory \cite{PAST}, which involves changes in the relative time delay of two freely-falling test masses with initial anti-orbital trajectories, and on the center-of-mass memory effect \cite{NICH1}, which refers to the changes in the relative time delay of two freely-falling test masses moving on anti-parallel trajectories initially. All these types of memories include the linear (ordinary) as well as the non-linear (null) contributions. The articles \cite{SERAJ1,  SERAJ2} discuss the idea of gyroscopic memory (precession of gyroscope leading to a net `orientation memory'). The permanent changes appearing in the background, i.e. the displacement, spin and center-of-mass memories, are related to the BMS (Bondi-Metzner-Sachs) transformations (eg. super-translations) and soft theorems \cite{STROM1, STROM2}.

The article \cite{IC1} suggests three different ways of deriving the memory effect, leading to qualitatively similar conclusions. The first procedure requires one to find a net displacement between a pair of geodesics, which appears due to the passage of an impulsive wave. This can be obtained simply by integrating the geodesic equations to determine the evolution of the separation between two nearby geodesics along each coordinate. The second method involves integrating the geodesic deviation equation to understand how the deviation vector evolves. The third way of arriving at gravitational memory is by analysing the behaviour of geodesic congruences. This covariant approach was proposed by O'Loughlin and Demirchian \cite{LOUGH}, who coined the term `$ B $-memory' in the context of impulsive GWs \cite{PEN2, STEIN3, PODOL, BATTIS, BHATT}.

Among the recent works, the extensive study of memory effects due to plane gravitational waves by Zhang and his collaborators is of interest \cite{ZHANG1, ZHANG2, ZHANG3, ZHANG7}. They have investigated linearly polarized exact plane waves with a Gaussian profile in \cite{ZHANG1, ZHANG2}, impulsive waves \cite{ZHANG3}, circularly polarized waves, and periodic waves in their subsequent papers. As mentioned in Ref.\cite{ZHANG1}, a `pulselike' profile required for solving the geodesic equations in a given spacetime can originate from plausible astrophysical sources. Memory effects in pp-wave spacetimes have also been studied in \cite{IC1, IC5, MALUF1, MALUF2, SIMIC1, SIMIC2, PREN}, where the wave profiles have been represented by Gaussian, sech-squared, square and other pulses.

In this paper, we assume a pulse profile in the shape of a ramp to mimic an impulse of GW generated e.g., during the merger of two black holes, and examine the memory effects these waves produce in a pp-wave spacetime. Such a pulse has not been investigated so far.
Solving for the geodesic equations in the absence as well as in the presence of a pulse will help us to understand how the separation of the geodesics evolves with time due to the passage of such a wave profile. We will also see whether the solutions can be analytically derived when a ramp profile is considered, as we already know that analytical solutions could not be found in the case of Gaussian profile \cite{ZHANG1,ZHANG2}, although it could be extracted for a particular case of Dirac-delta pulse \cite{PREN}, and for a square pulse \cite{IC1,IC5}.
Our paper is organised into the following sections. Sec.II presents the well-known pp-wave spacetime along with its characteristic features and the geodesic equations. In Sec.III, we discuss the physically viable pulse profiles which may be used to model GWs. We briefly review the memory effects produced by various profiles available in the literature, and then introduce our choice of pulse profile, which is the ramp waveform. We derive the analytical solutions of the geodesic equations both in the presence and absence of the pulse, and display the nature of memory in the respective plots for plus polarization in Sec.IV and for cross polarization in Sec.V. We conclude our study in Sec.VI with an analysis of our results and a comparison with the corresponding results obtained from geodesic deviation equations and those from similar works in the literature.

\section{Geodesics in pp-wave spacetimes}

The class of generalized pp-waves constitute one of the best-known and simplest set of solutions to the Einstein's field equations in General Relativity \cite{EHL}. The pp-waves were first studied by Brinkmann \cite{BRINK}, and interpreted as gravitational waves by Peres \cite{PERES}.
Their properties are elaborated in a number of articles (see e.g. \cite{EHL, GRIF, JORDAN, STEPH1}). The family of pp-wave spacetimes is characterised by a covariantly constant null vector field whose shear, expansion and twist are all zero. So they can be used as a model for gravitational or electromagnetic waves, or other forms of matter, or any combination of these.
The pp-waves are plane-fronted GWs with parallel rays. The rays being parallel, the rotation of the vector field vanishes. Scalar invariants of all orders derived from the corresponding Riemann tensor vanish. Therefore pp-waves are exact solutions of the full non-linear classical string theory \cite{HORO1, HORO2, SENO}, and can be applied in studying certain properties of GWs (e.g. their focusing properties and possible non-linear interactions), which cannot be explained by considering approximation schemes only \cite{GRIF}. The Petrov type of these spacetimes is N or O and the rank of the Riemann tensor is two. They belong to the wider Kundt's class of solutions exhibiting a  shear-free, twist-free, non-expanding null congruence \cite{STEPH1}.

In the conventional description of the pp-wave metric, the coordinates $ u $ and $ v $ are used instead of the standard Cartesian coordinates, where one defines \cite{MALUF1, MALUF2, ORTIN}:
\begin{equation}
u= \frac{1}{\sqrt{2}} (t-z), \quad \quad  v= \frac{1}{\sqrt{2}} (t+z).
\end{equation}
The covariantly constant null Killing vector field $ l_{\mu} $ admitted by pp-wave spacetime is related to the light-cone coordinates $ u $ and $ v $ by: $ l_{\mu} = \partial_{\mu} u $, and $ l^{\mu} \partial_{\mu} v =1 $. Therefore the metric can be made independent of only the $ v $-coordinate.
The Killing vector satisfies the relations: $ \nabla_{\mu} l_{\nu} =0 $, $ l^2 = l_{\mu} l^{\mu} =0 $. A gravitational wave propagating in the $(t, z)$-direction distorts the spacetime in transverse directions only. Hence it is advantageous to switch over to the light-cone coordinates. Moreover, by virtue of the property of admitting a covariantly constant null vector field, the pp-wave metric can be made independent of the $v$-coordinate.

The general pp-wave spacetimes include the plane wave spacetimes. Plane waves in Einstein-Maxwell theory were first considered by Baldwin and Jeffery \cite{BALD}.
The exact plane wave spacetimes are those where the Riemann curvature is constant over each wavefront.
It is known that every spacetime near a null geodesic can be approximated to a plane wave (Penrose limit). Discussions on Penrose limit and the memory effect are available in \cite{PEN1, HAW, MIS, BLAU, SHORE}.
One possible form \cite{EHL, JORDAN, STEPH2} of the plane-fronted gravitational wave travelling in the $ z $-direction is defined by the metric:
\begin{equation} \label{1}
ds^2 = - H(u,x,y) du^2 -2 du dv + dx^2 + dy^2.
\end{equation}

This metric is in the Brinkmann coordinate system, which is harmonic as well as global, free from coordinate singularities \cite{PERES, BONDI}. The Brinkmann coordinates are defined in terms of $(u,v,x,y)$ coordinates instead of the usual $(t,x,y,z)$ coordinates.
The quantity $ H(u,x,y) $ represents the profile and the polarization of the gravitational wave.
The presence of such a free function makes the metric advantageous for defining a pulse profile for the wave.
The corresponding vacuum Einstein equations lead to:
\begin{equation} \label{2a}
\partial_{x}^{2}H + \partial_{y}^{2}H =0,
\end{equation}
and its solution can be written as:
\begin{equation} \label{2b}
H(u,x,y) = A_{+}(u) \cdot \left(\frac{x^2 - y^2}{2}\right) + A_{\times}(u) \cdot xy.
\end{equation}
This solution satisfies the wave equation also. $ A_{+}(u) $ and $ A_{\times}(u) $ denote the `plus' and `cross' polarizations respectively.
Ehlers and Kundt \cite{EHL} have studied the vacuum pp-waves and determined all the possible forms of $ H $. The possible nature of $ H $, the choices of the polarization states and the respective Killing vector fields and symmetries
(including conformal symmetries) have been studied in \cite{SIPPEL, KUH, AICH, KEANE, HUSSAIN} as well.

The non-zero components of the Riemann tensor corresponding to the metric \eqref{1} are:
\begin{equation}\label{3}
R_{xuxu} = \frac{1}{2} A_{+}(u), \quad  R_{yuyu} = -\frac{1}{2} A_{+}(u), \quad
R_{xuyu} = \frac{1}{2} A_{\times}(u).
\end{equation}
Subsequently the geodesic equations are found to be:
\begin{equation}\label{5a}
\ddot{x}= -\frac{1}{2} A_{+} x -\frac{1}{2} A_{\times} y,
\end{equation}
\begin{equation}\label{5b}
\ddot{y}= \frac{1}{2} A_{+} y -\frac{1}{2} A_{\times} x,
\end{equation}
\begin{equation}\label{5c}
\ddot{v}= -\left[ \frac{1}{4}\dfrac{d A_{+}}{du} (x^2 - y^2) + \frac{1}{2}\dfrac{d A_{\times}}{du} xy +  A_{+} (x \dot{x} - y \dot{y}) + A_{\times} (x \dot{y} + y \dot{x}) \right].
\end{equation}
From the geodesic Lagrangian, the general form of $ \dot{v}(u) $ is obtained as \cite{IC5}:
\begin{equation}\label{5d}
\dot{v} = \frac{1}{2} (\dot{x}^2 + \dot{y}^2) - \frac{1}{4} A_{+} (x^2 - y^2) - \frac{1}{2} A_{\times} xy + \frac{k}{2}.
\end{equation}
On integration, this gives
\begin{equation}\label{5e}
v(u) = v_0 + \frac{1}{2} (x \dot{x} + y \dot{y} + ku),
\end{equation}
where $ v_0 $ is the integration constant, $ k $=0 and $1$ for null and time-like geodesics respectively,  $ u $ is an affine parameter, and the overdot denotes differentiation w.r.t. $ u $, throughout this paper.

Zhang \textit{et al.} \cite{ZHANG2} have presented a detailed discussion on the derivation of the geodesic equations and the physical significance of the related quantities. They have pointed out that the profile of the wave must be `pulselike', extending within a finite time span. This pulse-nature is encoded in the expressions for $ A_{+}(u) $ and $ A_{\times}(u) $.
If these expressions are known, one can determine each of the quantities: $ x $, $ y $, $ v $, $ \dot{x} $, $ \dot{y} $ and $ \dot{v} $ from the above equations. This is a conventional approach for investigating gravitational memories using different types of pulse profiles, which has been applied in earlier studies (\cite{ZHANG1, ZHANG2, ZHANG3, PREN, IC1, IC5, IC2, IC3}).

\section{Choices of pulse profiles}

\subsection{Pulse profiles used in earlier works}
Though simple in appearance, the above geodesic equations, in general, cannot be explicitly solved. Solutions can be obtained only in some particular cases, which are related to the symmetry of the metric. Those with the maximal symmetry present the most interesting cases \cite{PREN}. The isometry group of the plane GWs has the generic dimension of five \cite{EHL, JORDAN, SIPPEL}. It becomes six when $ A_{+}(u)= $ constant, or when $ A_{+} \propto u^{-2}$ \cite{PREN, EHL}.
The first wave profile can depict a gravitational wave sandwiched between two Minkowskian regions. The second profile is geodesically incomplete but useful in studying the Penrose limit.
However, in the non-flat case, with these two profiles, the pp-wave metric displays the maximal, 7-dimensional conformal symmetry (See \cite{PREN} and references therein). For plane GWs, there exists only one more metric family with 7 dimensions \cite{TUPPER, KUH}, described by  $ A_{+}(u) = \dfrac{c}{(u^2 +au +b)^2} $, where the denominator should be non-singular.
In Ref.\cite{PREN}, the pulse profile takes the particular form:
$ A_{+}(u) = \dfrac{2 \epsilon ^2}{\pi (u^2 +\epsilon ^2)^2} $,
and the geodesic equations can be solved analytically. Subsequently, the authors have explored how a classical particle interacts with the plane GWs. They have found that the energy of the particle after the passage of the pulse would be greater than, less than or equal to its initial energy, depending on the relations between the initial positions and velocities, which has been reported in \cite{MALUF1, MALUF2} also.

Among the pulse profiles considered for examining the gravitational memory effect, the Gaussian is the most widely studied. Zhang \textit{et al.} solved the geodesic equations numerically and analysed the memory effect produced by Gaussian pulses and their integrals and derivatives \cite{ZHANG1, ZHANG2}.
These profiles do not represent sandwich waves \cite{BONDI2, BONDI3} in the strictest sense as they are not confined within a small curved region between two flat spacetimes, but their curvatures vanish rapidly outside the pulse width. The first derivative of a Gaussian may mimic a wave pulse generated from a gravitational flyby, while the second and third derivatives may be related to the system presented in Ref.\cite{BRAG2} and to gravitational collapse respectively.
The authors have shown that the free particles move apart with a constant, non-zero velocity after the passage of the wave. They have also investigated the gravitational memory of impulsive waves \cite{ZHANG3} by squeezing a smooth Gaussian waveform to a Dirac-delta form. 
Here their results are similar to those obtained in the smooth case. In addition, the velocities of the particles initially at rest jump due to the pulse, and the trajectories are discontinuous along the (light-like) direction of the wave propagation.

Chakraborty and others \cite{IC1, IC5} have used the square pulse in their analysis of displacement and velocity memory effect. Such a profile depicts a gravitational wave burst, which occurs in a non-flat wave region sandwiched between two flat Minkowski spacetimes. The geodesics, which remain parallel before the arrival of the pulse, develop a finite, non-zero separation after the pulse disappears. The velocity difference between the geodesics also shows a sharp change in the wave region before attaining a non-zero final value. These changes can be summed up as monotonically increasing displacement memory and constant shift velocity memory along $ x $ and $ y $-directions \cite{IC5}. However, the solutions along the $ v $-direction are continuous but their derivatives are discontinuous at the boundaries of the pulse because of its step nature. Using square pulse and derivatives of sech-squared pulse, they have demonstrated that the kinematic variables such as expansion and shear carry information about memory \cite{IC1}. Generalising to the Kundt waves spacetimes, the authors have reported a new form of memory with sech-squared pulse in \cite{IC2}, as the geodesic separation exhibits periodic oscillations after the passage of the pulse.

From the above statements it is clear that both Gaussian pulses and square pulses can be actually encountered in astronomical considerations.

\subsection{Our choice of profile}
We choose a ramp waveform as the pulse profile in our study. The ramp waveform or a single sawtooth wave is one which rises linearly to its final value and then drops almost vertically. The ramp function is a commonly-used waveform that can be derived by integrating the Heaviside step function once or the Dirac-delta function twice. Before proceeding further, let us motivate our choice of pulse profile.

While determining memory effects from geodesic equation, the wave profile has to be `pulselike' \cite{ZHANG2}. The suitable choices are sandwich waves or shockwaves. We can consider a ramp profile as a sandwich wave since its curvature vanishes outside its width, and therefore as a suitable choice for representing the gravitational wave pulse.

The waveform models for GWs produced during binary black hole mergers have been constructed by Abbott \textit{et al.} (See e.g.\cite{ABBOTT1}) and several other authors in various studies. The wave strain versus time plots (as in Fig. 2 in \cite{LASKY}, Fig. 2 in \cite{ABBOTT1}, Fig. 6 in \cite{MILL}) indicate that the envelope of the waveform (i.e. wave amplitude $ h $),
rises slowly in the late inspiral phase and reaches a peak value at the time of the merger, before falling off rapidly during the ringdown phase. This nature of the wave envelope during inspiral and merger phases can be approximated by a ramp profile. Here we are looking at the peak values of the oscillations in the waveform (during inspiral and merger), so that the envelope can be visualized as a single ramp profile.

The advantage of this choice of pulse is that the geodesic equations can be solved analytically, as we shall show in the following sections, in contrast to the Gaussian pulses \cite{ZHANG1}, in which case analytical solutions are not available.
We shall also find that the nature of memory (velocity memory) appearing due to a ramp profile 
is similar to that obtained by Favata (Please see Fig. 1 in \cite{FAVATA})] who has considered the merger of inspiralling binaries. In view of that, we can mimic the merger phenomenon by a ramp waveform. It is known that the memory grows primarily in the final phase of merger \cite{MADI} in the case of gravitational wave bursts with memory.

Earlier studies on GW memory have considered pulse profiles representing step function (e.g. square pulses in \cite{IC1, IC5}) or profiles going over to the Dirac-delta function in the limiting case (e.g. Gaussians and their derivatives in \cite{ZHANG1, ZHANG2, MALUF1, MALUF2}, a particular pulse in \cite{PREN}). However the memory left behind in pp-wave spacetimes by a ramp profile has not been investigated so far.

\begin{figure}[h]
\begin{center}
\includegraphics[height=4cm,width=6cm]{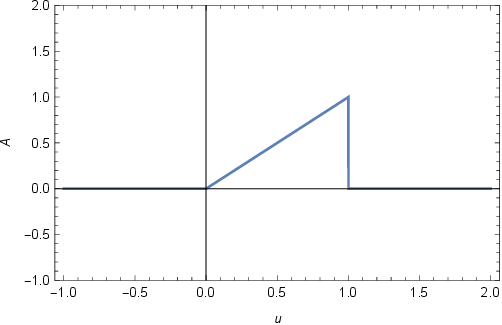}
\end{center}
\caption{A ramp profile of unit slope ($ \beta = 1 $) and unit width ($ a = 1 $).}
\label{fig:0}
\end{figure}

We consider a ramp profile of a finite width defined as:
\begin{equation}
\begin{split} \label{6a}
A_{\varepsilon}(u) = \beta u, \quad 0 \leq u \leq a,  \\
 =0 \qquad \text{elsewhere}.
\end{split}
\end{equation}
Here $ \varepsilon $ denotes $ + $ or $ \times $ polarizations, $ \beta $ is the slope and $ a $ is the width (and the peak value) of the pulse.
We aim to study how the passage of such a wave causes changes in the separation of two nearby geodesics and in their relative velocity. Any net non-zero differences in displacement and velocity will give us an idea of the memory effect produced by the GW pulse. A ramp of unit slope ($ \beta =1 $) and unit width ($ a =1 $) is shown in Fig. \ref{fig:0},  where $ A \equiv A_{\varepsilon} $ is plotted as a function of $ u $ for $ 0 \leq u \leq 1 $.

\section{Memory effects for plus polarization}
First, let us consider the memory effect due to plus polarization only. With $ A_{\times} =0 $, the geodesic equations \eqref{5a} and \eqref{5b} for the spatial coordinates $ x $ and $ y $ become
\begin{equation}
\ddot{x} =- \frac{1}{2} A_{+} x, \label{5f} \qquad \text{and} \qquad \ddot{y} = \frac{1}{2} A_{+} y. 
\end{equation}
From the Bargmann point of view \cite{ZHANG3, DUVAL1, DUVAL2}, these equations describe a non-relativistic particle subjected to a time-dependent anisotropic oscillator potential, which is attractive in the $ x $-coordinate and repulsive in the $ y $-coordinate.

If the explicit expression of the pulse profile $ A_{+}(u) $ is known, one can determine the solutions for $ x(u) $ and $ y(u) $, and subsequently $ v(u) $ using Eqn.\eqref{5e}. Plugging in the expression for $ A_{+}(u) $ from Eq.\eqref{6a}, equations in \eqref{5f} can be solved analytically, and we get
\begin{equation}\label{7a}
x(u) =
\begin{cases}
 k_1 u + k_2,  \hspace{7.5cm}  u \leq 0,  \\
 k_3 \cdot \text{Ai} \left( \left( -\dfrac{\beta}{2} \right)^{1/3} u \right)
 + k_4 \cdot \text{Bi} \left( \left( -\dfrac{\beta}{2} \right)^{1/3}  u \right),
\hspace{1cm} 0 \leq u \leq a, \\
 k_5 u + k_6,  \hspace{7.5cm}  a \leq u.
\end{cases}
\end{equation}

\begin{equation}\label{7b}
y(u) =
\begin{cases}
 l_1 u + l_2,  \hspace{7cm}  u \leq 0,  \\
 l_3 \cdot \text{Ai} \left( \left( \dfrac{\beta}{2} \right)^{1/3} u \right)
 + l_4 \cdot \text{Bi} \left( \left( \dfrac{\beta}{2} \right)^{1/3} u \right),
\hspace{1cm} 0 \leq u \leq a, \\
 l_5 u + l_6,  \hspace{7cm}  a \leq u.
\end{cases}
\end{equation}

Here $ k $'s and $ l $'s are the integration constants. Ai and Bi are the Airy functions.
Differentiating these w.r.t. $ u $, we have
\begin{equation}\label{7c}
\dot{x} (u) =
\begin{cases}
 k_1,  \hspace{10cm} u \leq 0,  \\
 \left( -\dfrac{\beta}{2} \right)^{1/3} \left\lbrace k_3 \dot{\text{Ai}} \left( \left( -\dfrac{\beta}{2} \right)^{1/3} u \right)
 + k_4  \dot{\text{Bi}} \left( \left( -\dfrac{\beta}{2} \right)^{1/3} u \right) \right\rbrace,
\hspace{0.8cm} 0 \leq u \leq a,  \\
 k_5,  \hspace{10cm} a \leq u.
\end{cases}
\end{equation}

\begin{equation}\label{7d}
\dot{y} (u) =
\begin{cases}
 l_1, \hspace*{9cm} u \leq 0, \vspace{0.1cm} \\
 \left( \dfrac{\beta}{2} \right)^{1/3} \left\lbrace l_3 \dot{\text{Ai}} \left( \left( \dfrac{\beta}{2} \right)^{1/3} u \right)
 + l_4  \dot{\text{Bi}} \left( \left( \dfrac{\beta}{2} \right)^{1/3} u \right) \right\rbrace,
\hspace{0.8cm} 0 \leq u \leq a,  \vspace{0.1cm} \\
 l_5,  \hspace{9cm}  a \leq u.
\end{cases}
\end{equation}

\begin{figure}[h]
\centering
\begin{minipage}[b]{0.3\linewidth}
\includegraphics[height=4cm,width=5.7cm]{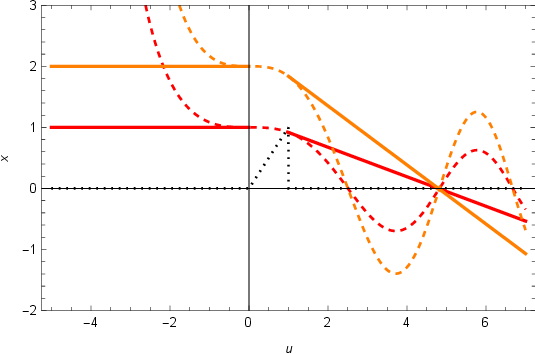}
\begin{center}
(a)
\end{center}
\end{minipage} \hspace{3cm}
\begin{minipage}[b]{0.3\linewidth}
\includegraphics[height=4cm,width=5.7cm]{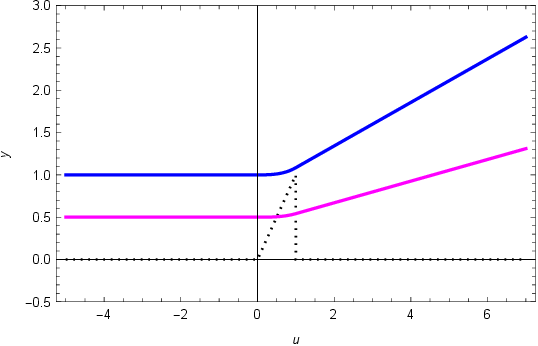}
\begin{center}
(b)
\end{center}
\end{minipage}
\caption{Displacement memory along $ x $ and $ y $-directions for plus polarization is shown for geodesics drawn assuming the values: $ k_2=1 $ (red), $ k_2=2 $ (orange), $ l_2=1 $ (blue), and $ l_2=0.5 $ (magenta). The dashed lines denote the Airy functions over the entire range of $ u $. In the region between $ u=0 $ and $ u=1 $, the required solutions are given by the dashed lines connecting the solid lines to the left of $ u=0 $ and to the right of $ u=1 $. This is done to avoid the limitations in plotting with Piecewise command in Mathematica. The black dotted lines represent the wave pulse.}
\label{fig:1}
\end{figure}

\begin{figure}[h]
\centering
\begin{minipage}[b]{0.4\linewidth}
\includegraphics[height=4cm,width=7cm]{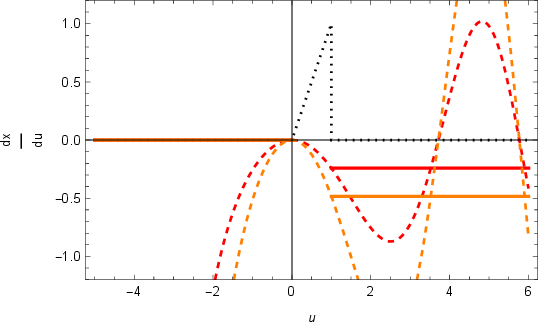}
\begin{center}
(a)
\end{center}
\end{minipage} \hspace{1.5cm}
\begin{minipage}[b]{0.4\linewidth}
\includegraphics[height=4cm,width=7cm]{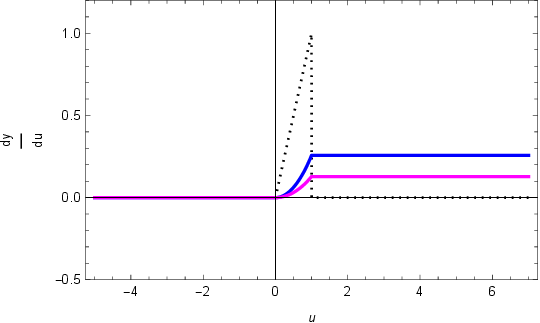}
\begin{center}
(b)
\end{center}
\end{minipage}
\caption{Velocity memory along $ x $ and $ y $-directions for plus polarization is shown for geodesics with $ k_2=1 $ (red), $ k_2=2 $ (orange), $ l_2=1 $ (blue), and $ l_2=0.5 $ (magenta).}
\label{fig:2}
\end{figure}

Here $ \dot{\text{Ai}} $ and $ \dot{\text{Bi}} $ represent the derivatives of Airy functions w.r.t. $ u $. The solutions in \eqref{7a} and \eqref{7b} denote the nature of a geodesic along $ x $ and $ y $-directions respectively. Solving for the constants, it is assumed that the separation between a pair of nearby geodesics remains constant initially, and hence the initial velocity will be zero. So we will have $ k_1=0 $.
As soon as the gravitational pulse arrives, the separation changes and the velocity of the changing separation becomes non-zero. This gives a measure of the displacement memory and velocity memory.
Assuming that the solutions of the geodesic equations are continuous and differentiable at the boundaries of the pulse, we equate the values of the functions $ x(u) $ and $ \dot{x} (u) $ at $ u=0 $ and $ u=1 $ (after choosing $ \beta = a=1 $), and  arrive at the following relations:
\begin{equation}
\begin{split}
k_3 = k_2/ 0.709,  \qquad  k_4 = 0.576 k_3, \hspace*{2cm} \\
k_5 = -0.149 k_3 + 0.575 k_4,  \qquad  k_6 = -k_5 + 0.171 k_3 + 1.308 k_4.
\end{split}
\end{equation}
We find that all the constants $ k_3 $, $ k_4 $, $ k_5 $ and $ k_6 $ can be evaluated from a single constant $ k_2 $, which is nothing but the initial value of $ x(u) $. Thus the memory effects are determined by the initial separation of the nearby geodesics. The values of the constants also depend on the points where the boundaries of the pulse are chosen, i.e. on the width and position of the pulse on the $ u $-axis.
Likewise along $ y $-direction, one can show that $ l_1 =0 $, and the value of $ l_2 $ determines the values of the remaining integration constants. Inserting the values of the integration constants, the solutions in \eqref{7a}-\eqref{7d} are plotted in Figs. \ref{fig:1} and \ref{fig:2}.


The curves in Figs. \eqref{fig:1}(a), (b), \eqref{fig:2}(a) and (b) respectively represent the quantities $ x(u) $, $ y(u) $, $ \dot{x}(u) $ and $ \dot{y}(u) $ with two sets of initial values. The pulse in the shape of a ramp is outlined by the black dotted lines. The red curves are obtained with $ k_2=1 $, the orange ones with $ k_2=2 $, the blue curves with $ l_2=1 $, and the magenta ones with $ l_2=0.5 $. In Figs. \eqref{fig:1}(a) and \eqref{fig:2}(a), the dashed lines denote the Airy functions (from Eqs.\eqref{7a} and \eqref{7c}) over the entire range of $ u $. In the region between $ u=0 $ and $ u=1 $, the dashed lines connecting the solid lines to the left of $ u=0 $ and to the right of $ u=1 $ show the required solutions.
The plots in \eqref{fig:1}(b) and \eqref{fig:2}(b) respectively represent the evolution of the separation of two nearby geodesics and their velocity profiles along $ y $-direction. The separation, which is initially constant, monotonically increases after the arrival of the pulse and leads to a net displacement. On the other hand, the geodesics are co-moving initially but their relative velocity shows a rise while the pulse lasts, and reaches a constant value after it leaves. Thus a non-zero relative velocity is left behind as memory (constant shift memory \cite{IC5}). Similar nature is observed in the corresponding plots for $ x $-direction.

We have found that even if we change the slope $\beta$, e.g., $\beta=2 \, \textrm{or} \, 5$, and the width of the pulse $a$ ranging from $0$ to $2$, there is no significant difference in the nature of the plots.

\section{Memory effects for cross polarization}
In case of cross polarization, with $ A_{+} =0 $, the geodesic equations \eqref{5a} and \eqref{5b} are read as
\begin{equation}
\ddot{x}= -\frac{1}{2} A_{\times} y, \label{5g} \qquad \text{and} \qquad \ddot{y}= -\frac{1}{2} A_{\times} x. 
\end{equation}
We are going to use the ramp waveform \eqref{6a}, same as that in plus polarization, to denote the pulse profile of $ A_{\times} $. Introducing normal coordinates: $ p(u) = \dfrac{x+y}{2} $ and $ q(u) = \dfrac{x-y}{2} $, it can be shown that $ p(u) $ and $ q(u) $ satisfy equations similar to Eqs.\eqref{5f}. So the respective solutions will be similar. Reverting to the old coordinates $ x $  and $ y $, we can write the solutions of Eqs.\eqref{5g} as
\begin{equation}\label{8a}
x(u) =
\begin{cases}
\dfrac{1}{2} (mk_2 + nl_2),  \hspace{8cm}  u \leq 0,  \vspace{0.1cm} \\
\dfrac{1}{2} \left[
m\left\lbrace k_3 \cdot \text{Ai} \left( \left( -\dfrac{b}{2} \right)^{1/3} u \right)
+ k_4 \cdot \text{Bi} \left( \left( -\dfrac{b}{2} \right)^{1/3} u \right) \right\rbrace
\right.  \\  \left.
+ n\left\lbrace l_3 \cdot \text{Ai} \left( \left( \dfrac{b}{2} \right)^{1/3} u \right)
+ l_4 \cdot \text{AiryBi} \left( \left( \dfrac{b}{2} \right)^{1/3} u \right) \right\rbrace
\right],   \hspace{1cm}  0 \leq u \leq a,  \vspace{0.1cm} \\
\dfrac{1}{2} \left[ m(k_5 u + k_6) + n(l_5 u + l_6) \right],  \hspace{5.5cm}  a \leq u.
\end{cases}
\end{equation}

\begin{equation}\label{8b}
y(u) =
\begin{cases}
\dfrac{1}{2} (mk_2 - nl_2),  \hspace{8cm}  u \leq 0,  \vspace{0.1cm} \\
\dfrac{1}{2} \left[
m\left\lbrace k_3 \cdot \text{Ai} \left( \left( -\dfrac{b}{2} \right)^{1/3} u \right)
+ k_4 \cdot \text{Bi} \left( \left( -\dfrac{b}{2} \right)^{1/3} u \right) \right\rbrace
\right.  \\  \left.
- n\left\lbrace l_3 \cdot \text{Ai} \left( \left( -\dfrac{b}{2} \right)^{1/3} u \right)
+ l_4 \cdot \text{Bi} \left( \left( -\dfrac{b}{2} \right)^{1/3} u \right) \right\rbrace
\right],   \hspace{1cm}  0 \leq u \leq a,  \vspace{0.1cm} \\
\dfrac{1}{2} \left[ m(k_5 u + k_6) - n(l_5 u + l_6) \right],  \hspace{5.5cm}  a \leq u.
\end{cases}
\end{equation}

Subsequently, differentiating these w.r.t. $ u $, we have
\begin{equation}\label{8c}
\dot{x} (u) =
\begin{cases}
 0, \hspace*{10.5cm} u \leq 0, \vspace{0.1cm} \\
\dfrac{1}{2} \left[
m \left( -\dfrac{\beta}{2} \right)^{1/3} \left\lbrace k_3 \dot{\text{Ai}} \left( \left( -\dfrac{\beta}{2} \right)^{1/3} u \right)
 + k_4 \dot{\text{Bi}} \left( \left( -\dfrac{\beta}{2} \right)^{1/3} u \right) \right\rbrace
\right.  \\  \left. 
+ n \left( \dfrac{\beta}{2} \right)^{1/3}  \left\lbrace  l_3 \dot{\text{Ai}} \left( \left( \dfrac{\beta}{2} \right)^{1/3} u \right)
 + l_4 \dot{\text{Bi}} \left( \left( \dfrac{\beta}{2} \right)^{1/3} u \right) \right\rbrace
\right],
\hspace{1.4cm}  0 \leq u \leq a,  \vspace{0.3cm} \\
 \dfrac{1}{2} \left[ mk_5 u + nl_5 u \right],  \hspace{8cm}  a \leq u.
\end{cases}
\end{equation}

\begin{equation}\label{8d}
\dot{y} (u) =
\begin{cases}
 0, \hspace*{10.5cm} u \leq 0, \vspace{0.1cm} \\
\dfrac{1}{2} \left[
m \left( -\dfrac{\beta}{2} \right)^{1/3} \left\lbrace  k_3 \dot{\text{Ai}} \left( \left( -\dfrac{\beta}{2} \right)^{1/3} u \right)
 + k_4 \dot{\text{Bi}} \left( \left( -\dfrac{\beta}{2} \right)^{1/3} u \right) \right\rbrace
\right.  \\  \left.
- n \left( \dfrac{\beta}{2} \right)^{1/3} \left\lbrace  l_3 \dot{\text{Ai}} \left( \left( \dfrac{\beta}{2} \right)^{1/3} u \right)
 + l_4 \dot{\text{Bi}} \left( \left( \dfrac{\beta}{2} \right)^{1/3} u \right) \right\rbrace
\right],
\hspace{1.4cm}  0 \leq u \leq a,  \vspace{0.3cm} \\
 \dfrac{1}{2} \left[ mk_5 u - nl_5 u \right],  \hspace{8cm}  a \leq u.
\end{cases}
\end{equation}

These solutions represent the behaviour of geodesics and their relative velocities along $ x $ and $ y $-directions. Here the values of $ m $ and $ n $ can be chosen arbitrarily. The derivatives of Airy functions are shown by the overdots on Ai and Bi. $ k $'s and $ l $'s are the integration constants to be determined in the same way as done in case of plus polarization. Analogous to what we have found in the previous section, we will have here $ k_1 = l_1 = 0 $, and hence we have omitted $ k_1 $ and $ l_1 $ in the solutions for $ u \leq 0 $ in \eqref{8c} and \eqref{8d}). By setting the values of $ k_2 $ and $ l_2 $ only, we can calculate the remaining integration constants. The nature of the above solutions can be determined from the following plots.

\begin{figure}[h]
\centering
\begin{minipage}[b]{0.3\linewidth}
\includegraphics[height=4cm,width=6cm]{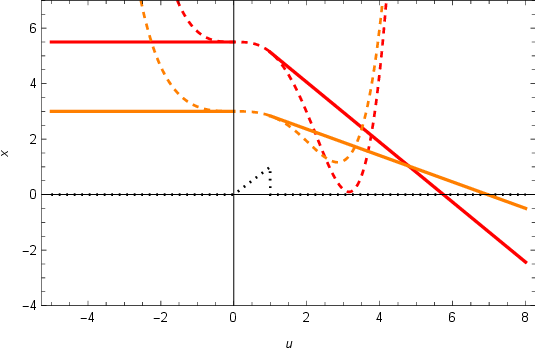}
\begin{center} (a)
\end{center}
\end{minipage} \hspace{3cm}
\begin{minipage}[b]{0.3\linewidth}
\includegraphics[height=4cm,width=6cm]{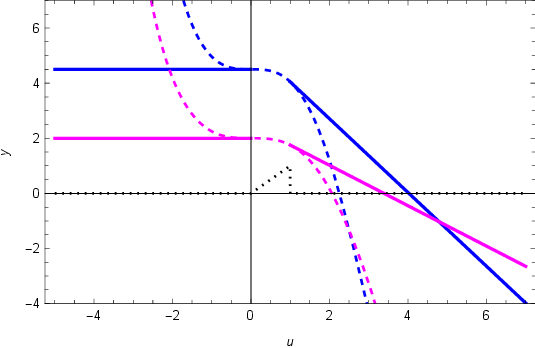}
\begin{center} (b)
\end{center}
\end{minipage}
\caption{Displacement memory along $ x $ and $ y $-directions for cross polarization is shown for geodesics drawn with $ k_2=5, l_2=1, m=2, n=1 $ (red), $ k_2=5, l_2=1, m=1, n=1 $ (orange), $ k_2=5, l_2=1, m=2, n=1 $ (blue), and $ k_2=5, l_2=1, m=1, n=1 $ (magenta). The dashed lines denote the Airy functions over the entire range of $ u $. In the region between $ u=0 $ and $ u=1 $, the dashed lines connecting the solid lines to the left of $ u=0 $ and to the right of $ u=1 $ represent the required solutions.  The black dotted lines outline the wave pulse.}
\label{fig:3}
\end{figure}
\begin{figure}[h]
\centering
\begin{minipage}[b]{0.4\linewidth}
\includegraphics[height=4cm,width=7cm]{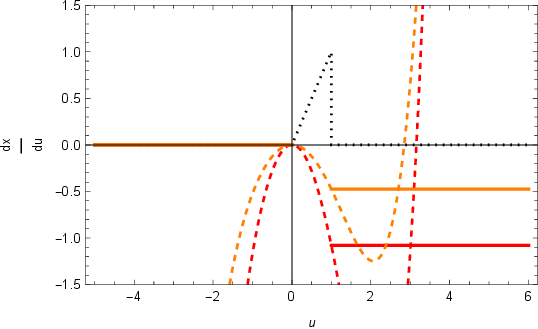}
\begin{center} (a)
\end{center}
\end{minipage} \hspace{1.5cm}
\begin{minipage}[b]{0.4\linewidth}
\includegraphics[height=4cm,width=7cm]{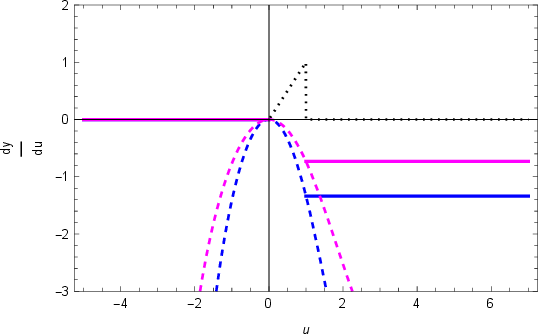}
\begin{center} (b)
\end{center}
\end{minipage}
\caption{Velocity memory along $ x $ and $ y $-directions for cross polarization is shown for geodesics drawn with $ k_2=5, l_2=1, m=2, n=1 $ (red), $ k_2=5, l_2=1, m=1, n=1 $ (orange), $ k_2=5, l_2=1, m=2, n=1 $ (blue), and $ k_2=5, l_2=1, m=1, n=1 $ (magenta).}
\label{fig:4}
\end{figure}

The curves in Figs. \eqref{fig:3}(a), (b), \eqref{fig:4}(a) and (b) respectively represent $ x(u) $, $ y(u) $, $ \dot{x}(u) $ and $ \dot{y}(u) $ where two sets of initial values have been chosen. The red curves are drawn with $ k_2=5, l_2=1, m=2, n=1 $, the orange ones with $ k_2=5, l_2=1, m=1, n=1 $, the blue curves with $ k_2=5, l_2=1, m=2, n=1 $, and the magenta ones with $ k_2=5, l_2=1, m=1, n=1 $. Just as in case of plus polarization, here also the dashed lines denote the Airy functions (from Eqs.\eqref{8a}-\eqref{8d}) over the entire range of $ u $. In the region between $ u=0 $ and $ u=1 $, the required solutions are given by the dashed lines connecting the solid lines to the left of $ u=0 $ and to the right of $ u=1 $.

The plots in \eqref{fig:3}(a) and (b) indicate that there is monotonically increasing displacement memory along both $ x $ and $ y $-directions. Figs. \eqref{fig:4}(a) and (b) show that constant shift velocity memory appears for both directions. The geodesic solutions for cross polarisation are nothing but the combinations of the $x$ and $y$ solutions for plus polarisation. Thus the corresponding effects are found to be similar in plus and cross polarizations.

\section{Discussions}

In this paper we have used a conventional method of analysing the memory effects of gravitational waves. The geodesic equations are solved in the presence of the wave as well as in the region beyond the extent of the wave in order to find the change in the separation between a pair of geodesics. The net changes in the displacement and velocity difference for two initially co-moving geodesics are interpreted as the gravitational memory effect.

We have considered the pp-wave metric in the Brinkmann coordinates. The metric contains a free function $ H (u,x,y) $ which can be so chosen as to denote a particular profile of the wave pulse. We have taken the pulse in the shape of a ramp to model burst GWs and investigated the memory it leaves behind in a pp-wave spacetime.
By integrating the geodesic equations in the presence of a ramp profile, we have derived the analytical solutions for $ x(u) $ and $ y(u) $ in terms of Airy functions. The solutions obtained in presence of the pulse are matched with those outside the wave region, based on the assumption of the continuity and differentiability of the solutions at the boundaries of the pulse (i.e. $ u=0 $ and $ u=1 $). The initial values of $ x $, $ y $ and their first-order derivatives are chosen. The integration constants are then determined from the initial and boundary conditions.

For the sake of convenience in our calculations, we have assumed $ \beta =1 $, and $ a=1 $, for the pulse confined in the region between $ u=0 $ and $ u=1 $. From the set of figures shown in the previous sections (Secs. IV and V), one can find the changes in the geodesic separation and velocity profiles as the pulse hits. The separation initially remains constant for the two geodesics. When the pulse arrives, lasts for a short duration and finally dies out, the geodesic separation indicating the displacement memory goes on increasing monotonically. This holds for both $ x $ and $ y $-directions irrespective of the plus and cross polarizations. On the other hand, the velocity memory effects given by the derivatives of the separation along $ x $ and $ y $-directions increase from the initial zero value because of the pulse, before reaching a constant non-zero value after the decay of the pulse. Thus a constant shift velocity memory remains after the wave ceases to exist. The same behaviour is observed for both types of polarizations. Our results agree with those obtained in pp-wave spacetime due to the passage of GW pulses represented by some other profiles: Gaussian pulse in \cite{ZHANG1, ZHANG2} and square pulse in \cite{IC1, IC5}. In plus polarization, the changes in displacement and velocity along $ x $-direction display opposite orientations w.r.t. to those $ y $- directions because of the opposite signs appearing in the geodesic equations in \eqref{5f}. The combinations of the geodesic solutions for plus polarization are taken as the solutions for cross polarization, and the net effects of such combinations yield the plots where the changes in $ x $, $ y $, $ \dot{x} $ and $ \dot{y} $ show the same sign. But the physical interpretation of the nature of changes remains the same in all cases.  

Analytical solutions to geodesic equations have been extracted for a particular case of Dirac-delta pulse by Andrzejewski and Prencel \cite{PREN} and for a square pulse by Chakraborty and Kar \cite{IC1, IC5}. The respective solutions are found to be combinations of inverse trigonometric functions, and of sinusoidal and hyperbolic functions. In contrast, our solutions for a ramp profile appear as Airy functions. Therefore we can say that the nature of the solutions depends on the shape of the pulse profile. It is interesting to note that although the analytical solutions are different for different types of profiles, but the overall nature of the memory effect is very much similar. The similarity may be due to the fact that all these studies have assumed a pp-wave spacetime in Brinkmann coordinates with initially comoving geodesics, and examined the effect of only an isolated pulse. Compared to the single square pulse \cite{IC1, IC5}, the single ramp profile is more suitable as a model of GW pulse. This profile is important in the sense that we have been able to investigate it analytically and reproduce the results obtained in \cite{IC1, IC5}.

Changing the slope or width of the pulse or its location on the positive $ u $-axis will change the boundary values of $ x(u) $, $ y(u) $, $ \dot{x}(u) $ and $ \dot{y}(u) $, and hence the values of the integration constants appearing in the solutions. But that the behaviour of the plots is found to remain unchanged in each case.
For the changes along the $ v $-direction, we determine the expressions which are too complicated to be shown graphically.

To confirm our results we looked into the geodesic deviation equation to examine the memory effects. Considering two infinitesimally close geodesics, $ X_1^{\mu} $ and $ X_2^{\mu} = X_1^{\mu} + \eta^{\mu} $, we have the geodesic deviation equation given by:
\begin{equation}
\frac{d^2 \eta^{\mu}}{d \tau ^2} + R^{\mu}_{\alpha \nu \beta} \frac{dX^{\alpha}}{d\tau} \frac{dX^{\beta}}{d\tau} \eta^{\nu} =0,
\end{equation}
where $ \frac{dX^{\alpha}}{d\tau} $ is the unit tangent vector and $ \eta^{\nu} $ is the connecting vector (See Sec.III.A of \cite{ZHANG2} and references therein). Putting $ u= \tau $, $ v= -\frac{1}{2} \tau $, $ X^1= X^2= 0 $, this equation reduces to
\begin{equation} \label{9a}
\frac{d^2 \eta^{1}}{du^2} = -\frac{1}{2} A_{+} \eta^{1} -\frac{1}{2} A_{\times} \eta^{2}, \qquad
\frac{d^2 \eta^{2}}{du^2} = \frac{1}{2} A_{+} \eta^{2} -\frac{1}{2} A_{\times} \eta^{1}, \qquad
\frac{d^2 \eta^{3}}{du^2} =0.
\end{equation}

These equations can be re-derived \cite{ZHANG2} from the geodesic equations (Eqs.\eqref{5a}-\eqref{5c}) corresponding to the line element \eqref{1}. With reference to a Fermi coordinate system ($ x^0 $, $ x^i $), (where the metric is locally flat, $ x^0 $ coincides with the proper time $ \tau $ along the geodesic and the coordinate is at rest with respect to a freely falling detector), the acceleration of the geodesic separation $ \eta^i $ experiences a forcing term $ -R^{i}_{0j0} \eta^{j} $ \cite{ZHANG2}. This forcing term in the linearized theory, as mentioned by Zhang \textit{et al.}, would be `pulselike'. Since the change in separation $ x^i = \eta^i - l^i $ as well as the curvature are small, the geodesic deviation can be approximated as $ \dfrac{d^2 x^{i}}{dt^2} = -R^{i}_{0j0} l^{j}. $
Here $ l^i $is the time-averaged separation. Also, it is known that in the linear theory, $ R_{i0j0}= \dfrac{G}{3r} \dfrac{d^4 D_{ij}}{dt^4} (t-r) $, where $ D_{ij} $ is the quadrupole moment of the source at distance $ r $, with the retarded time $ u= t-r $.

For plus polarization only (i.e. $ A_{\times} =0 $), the geodesic deviations along $ x $ and $ y $-directions (from Eqs.\eqref{9a}) satisfy the equations:
\begin{equation} \label{9b}
\frac{d^2 \eta^{x}}{du^2} = -\frac{1}{2} A_{+} \eta^{x}, \qquad
\frac{d^2 \eta^{y}}{du^2} = \frac{1}{2} A_{+} \eta^{y},
\end{equation}
which are similar to the geodesic equations \eqref{5a} and \eqref{5b}. The deviation vectors corresponding to the two geodesics $ (x_1, y_1) $ and $ (x_2, y_2) $ will be obtained as
\begin{equation}
\eta^{x}= x_1- x_2 =x_1, \qquad \eta^{y}= y_1- y_2 =y_1
\end{equation}
when we assume $ x_2 = y_2 =0 $.

So, by using the geodesic deviation equations, we arrive at the results identical to those derived from the geodesic equations, thereby validating our results. This is expected because the pp-wave line element describes a flat spacetime over which the curvature perturbations travel. If the background spacetime itself has non-zero curvature, it will contribute, in addition to the wave, to the total deviation. In such case, one has to consider Fermi normal coordinates and obtain the geodesic deviation in tetrad basis, which can be split into the background and the wave parts \cite{IC3}. The total deviation is found to yield qualitatively similar results on memory to those determined from the geodesics.

In the context of detection of the gravitational memory effect, we note that different gravitational waves when passing through a medium are expected to leave different permanent changes that will lead to chaos of the gravitational wave memory. In order to detect the permanent memory belonging to a specific event of generation of a gravitational wave, we require improved detection techniques in the next generation detectors. The memory component is only a small fraction of the total strain produced during a merger phenomenon \cite{Favata1}. So it is highly unlikely that the Advanced LIGO would detect the memory corresponding to a specific coalescence signal. Till ALIGO reaches the desired design sensitivity, it is possible to incorporate higher order gravitational wave modes for parameter estimation of binary black hole mergers, and these methods can also be used to detect higher-order modes themselves, as was proposed by Lasky \textit{et al.} \cite{LASKY}.
The continued detections of gravitational waves from binary black hole mergers and binary neutron star inspirals have opened up a new era of gravitational wave astronomy. In addition to the isolated events, there is a host of weak, unresolved sources at higher redshifts, giving rise to the stochastic background which may be detected by the Advanced LIGO and Virgo detectors when the desired sensitivity is achieved by these detectors. There will also be superposition from other unresolved sources. The increased sensitivity of the third generation gravitational-wave detectors (Einstein Telescope and Cosmic Explorer) will make it possible to resolve and detect most of the binary black hole mergers, even the ones at high redshifts. Such considerations have led to the works on Compact Binary Foreground Subtraction to detect the gravitational wave background by the next generation detectors, by Regimbau \textit{et al.} \cite{Regimbau} and followed by Sachdev \textit{et al.} \cite{Sachdev}, Zhou \textit{et al.} \cite{Zhou}, Song \textit{et al.} \cite{Song}, and other similar workers.

We can consider another aspect of gravitational memory -- the change in kinetic energy of the free particle due to the passage of the wave. It has been found in \cite{MALUF1, MALUF2, PREN} that after the pulse dies out, the energy of the particle increases or decreases, depending on the initial conditions (position and velocity of the particle). The Newtonian work-energy relation is found to hold for a particle moving along geodesics in the pp-wave spacetimes, associated with a variation of the gravitational potential energy. However, the variation of the kinetic energy of free particles is not a useful criterion for investigating the memory effect, since both kinds of variation occur for the same wave. An exchange of energy takes place locally between the particle and the gravitational field of the wave \cite{MALUF3}. These results would be valid for our case too, where the wave pulse is modelled by a ramp profile. Inserting suitable values of initial positions and velocities, we can show that the kinetic energy of the particle either increases or decreases due to interaction with the gravitational wave. In the analysis of GW memory from geodesic equations, the test particles are considered as `idealized entities' that do not affect the gravitational field of the wave \cite{MALUF3}.  Unlike massive bodies like black holes, which produce gravitational field and whose motion gives rise to gravitational waves, the test particles can act only as probes of the field they are residing in. Hence, the accelerated test particle will radiate energy and its speed will decrease, but the emitted energy will not be in the form of gravitational wave. The particles are free before and after the passage of the GW.

In his paper, Kip Thorne \cite{THORNE} introduced a unified notation for the multipole formalism of gravitational waves. The energy, linear momentum, and the angular momentum in the waves were presented as infinite sums of multipole contributions. The Newtonian, post-Newtonian and post-post Newtonian formula for the multipole moments of slow-motion systems with weak internal gravity, were also provided. Among other aspects dealt with in his paper, the structure of the general solution of Einstein field equations was derived in terms of a sum of products of multipole contributions, in the full nonlinear general relativity. Subsequently, several authors studied the post-Newtonian correction to the quadrupole equations \cite{BD1}, even beyond the quadrupole approximation \cite{IW} and in the higher order \cite{BDE-FI}, as well as the spin multipole moment effects \cite{DI, Kidder}.

In the generation of GWs, it is the difference in the initial and final quadrupole moments of the source which gives rise to the gravitational wave memory \cite{BRAG1}. Non-linear memory appears for GWs which originate due to the mass-energy distribution of linear GWs. When linear GWs back-scatter on the spacetime curvature induced by the total mass-energy of the source, GW tails are generated. These phenomena occur due to quadratic multipolar interactions (mass-quadrupole and quadrupole-quadrupole interactions) \cite{Trestini}. The tail effect has been detected by the LIGO-Virgo collaboration \cite{ABBOTT2}. In recent times, many modified theories of gravity have been proposed to overcome the finite regime of validity of General Relativity. Some articles  have discussed the contribution of additional terms (boundary terms \cite{EXT1}, terms dependent on position, velocity, acceleration, and changing density of source \cite{EXT2}) in linearized theory to the generation of gravitational waves. Especially, the case of supernova explosion, involving a change in density and the case of two colliding meteorites may be addressed more accurately, if these additional terms are considered. Taking these into account, investigation of the gravitational memory effect in the modified theories of gravity will be of considerable interest in future.

\section*{Acknowledgement}
SD acknowledges the financial support from INSPIRE (AORC), DST, Govt. of India (IF180008). SG thanks IUCAA, India for an associateship.

\section*{Data Availability}
Data sharing is not applicable to this article as no datasets were generated or analysed during the current study. All figures were plotted using the theoretical equations.

\end{document}